# New low-temperature phases in $Na_xCoO_2$ linked to quantum criticality

F. Rivadulla[1*], M. Bañobre-López[1], M. García-Hernández[3], M. A. López-Quintela[1], J. Rivas[2]

[1]*Physical-Chemistry and* [2]*Applied-Physics Departments, University of Santiago de Compostela, 15782-Santiago de Compostela, Spain.*
[3]*Instituto de Ciencia de Materiales, CSIC,Campus de Cantoblanco 28049-Madrid.*

New electronic phases have been identified and placed in the (T,H) phase diagram of metallic $Na_xCoO_2$. At low Na-content (x ≈ 0.36), the magnetic susceptibility diverges with a power law $T^{-\eta}$ ($\eta$<1) and shows (T,H) scaling, indicating the proximity to a magnetic quantum phase transition. At high Na contents (x ≈ 0.6) the mass of the quasiparticles does never diverge, but renormalizes and becomes strongly field dependent at low temperatures, forming a heavy Fermi-Liquid. Our results make superconducting $Na_xCoO_2$ a clear candidate for magnetically mediated pairing.

It is commonly believed that the microscopic explanation for high-temperature superconductivity can be constrained by determining the interplay between band-filling, magnetic interactions and dimensionality. In this sense, the report of superconductivity in $Na_xCoO_2$•$yH_2O$[1] generated an intense debate about the nature of the pairing interaction in this material, which is becoming a test ground for the theories of superconductivity. Although the layered structure, the mixed valence character, and the proximity of the superconductive compositions to a localized charge-ordered phase[2] makes it very tempting to make a straightforward comparison with the cuprates, the metallic precursor of the superconducting phase shows its own peculiarities: an electronically active triangular cobalt lattice is commonly believed to be responsible for the strong temperature dependence of the metallic susceptibility (so called "Curie-Weiss metal" behavior), the anomalous temperature dependence of the resistivity,[3] or the surprisingly large thermopower.[4] Moreover, LDA calculations[5] predicted a ferromagnetic (FM) ground state in the metallic precursor, although the material never orders down to the lowest temperature probed. The discrepancy between theory and experiment was tentatively attributed to the proximity to a quantum phase transition (QPT), which would imply the existence of strong FM fluctuations in the metal. This opens the possibility of a magnetically mediated superconducting state due to exchange of FM spin fluctuations, like it was proposed for high-pressure $UGe_2$[6]. Actually, $^{59}$Co NMR and $^{23}$Na NQR experiments[7] revealed an unconventional form of a superconducting spin-triplet phase in $Na_xCoO_2 yH_2O$, as well as the existence of strong FM spin fluctuations in the metallic counterpart, confirmed by inelastic neutron scattering.[8]

Having in mind that understanding superconductivity requires a deep knowledge of the electronic precursor from which it forms, we have performed a detailed magnetic study of metallic $Na_xCoO_2$ at representative compositions of the series. For high-Na content ($x \approx 0.6$, when superconductivity is never achieved under hydration) we have found that the effective mass of the quasiparticles, m*, in the Fermi liquid (FL) state is renormalized at low temperature, forming a *heavy*-FL below ≈ 4 K. On the other hand, at $x \approx 0.36$, near the optimum value for superconductivity, strong magnetic fluctuations introduce a new phase with non-FL behavior and spin-glass relaxation. Application of a magnetic field restores the FL state of the metal in both cases, with m* strongly field dependent. A detailed scaling analysis of the singular low temperature $\chi(T,H)$ suggests that these effects are due to the proximity of metallic $Na_xCoO_2$ to a magnetic phase transition at T=0.

Polycrystalline samples of $Na_xCoO_2$ ($0.33 \leq x \leq 0.69$) were prepared by solid state reaction and subsequent chemical deintercalation of Na from the x=0.69 phase. The Na and Co contents were determined by inductively coupled plasma optical emission spectroscopy (ICP-OES), and the oxidation state of cobalt was determined by iodometric titration. Superconducting samples were obtained by stirring $Na_xCoO_2$, x < 0.45, in water for two days at room temperature. All the results discussed here correspond to single-phase materials; we paid special attention to avoid any trace of $Co_3O_4$ in our samples. Although we observed similar divergent susceptibility in single crystals (and it was observed in independent studies in the literature, see for example ref. [9,10,3]), we decided to use polycrystalline samples for this work due to the better bulk homogeneity after the Na-deintercalation process. In any case, the results presented here for low x-materials correspond to samples that we have verified to become superconductive after water insertion.

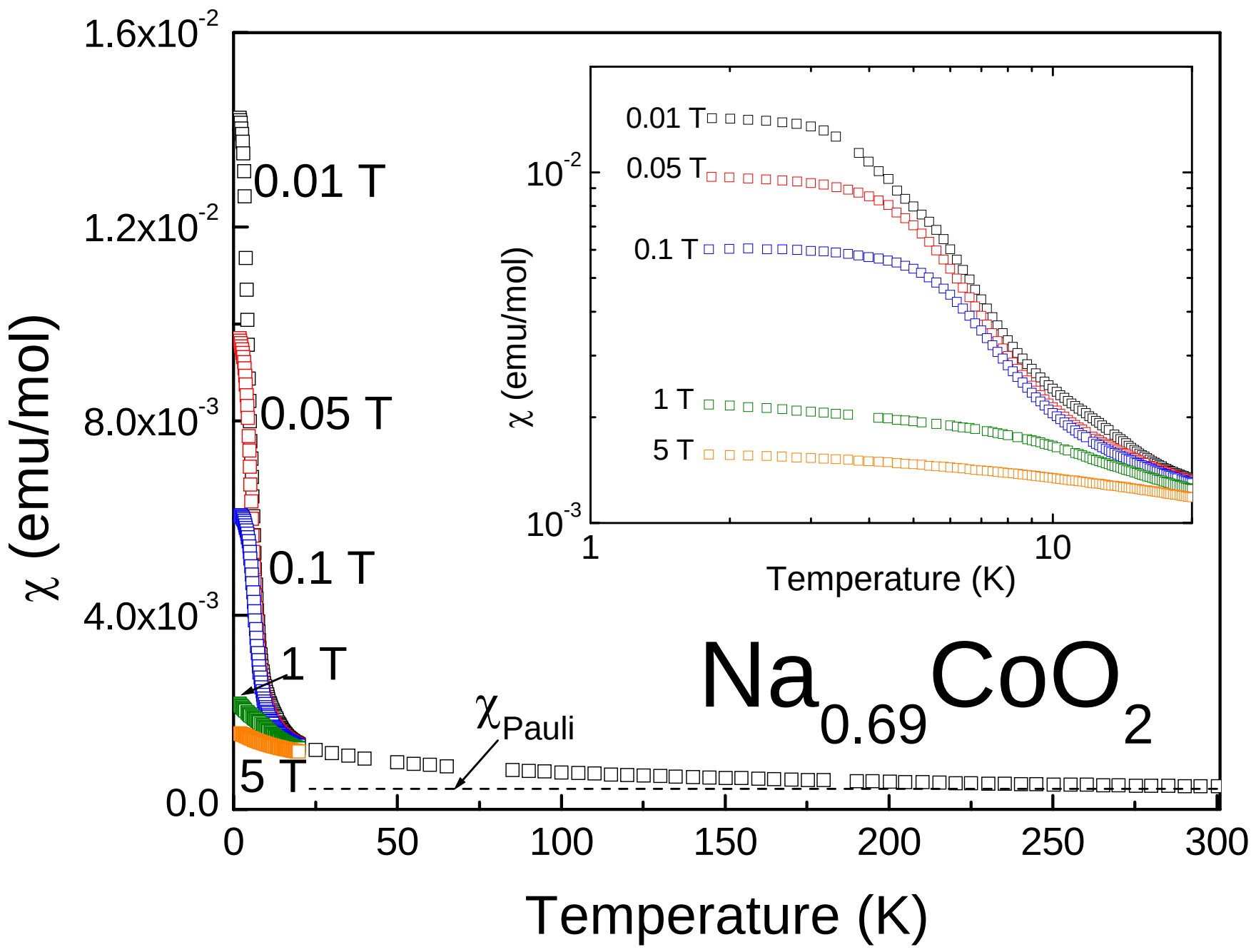

Fig.1.- Temperature and field dependence of the zero-field cooling (ZFC) magnetic susceptibility in the parent phase (x=0.69). The dotted line is the Pauli susceptibility determined from high temperature M(H) isotherms. Inset: Logarithmic plot of the low temperature susceptibility.

The temperature and field dependence of the magnetic susceptibility is shown in Fig. 1 for the parent phase (x=0.69). Low-field $\chi(T)$ shows a large enhancement over the free electron value below ≈10 K. First of all, we have considered the possibility of a paramagnetic impurity as the origin of the χ(T) divergence. However, the M(H) isotherms show strong deviations from the M vs. H/T scaling expected for the curvature of the Brillouin function at low temperatures and high fields (not shown). On the other hand, the departure of χ(T) from a temperature-independent Pauli paramagnetism, is strongly suppressed by a magnetic field, which suggest that the enhancement is due to the presence of magnetic fluctuations. A closer inspection (inset of Fig. 1) reveals that χ(T) flattens and becomes constant below ≈4 K. The increase in the constant $\chi(T\rightarrow 0)$ value is sensing the renormalization of the m* of the quasiparticles at the Fermi surface ($m^* \propto H^x$, $x \approx -0.4$). In this case, multiparticle correlations renormalize the FL parameters, but remain short-ranged, keeping the quasiparticles strictly independent in momentum space, even in the presence of strong magnetic fluctuations.

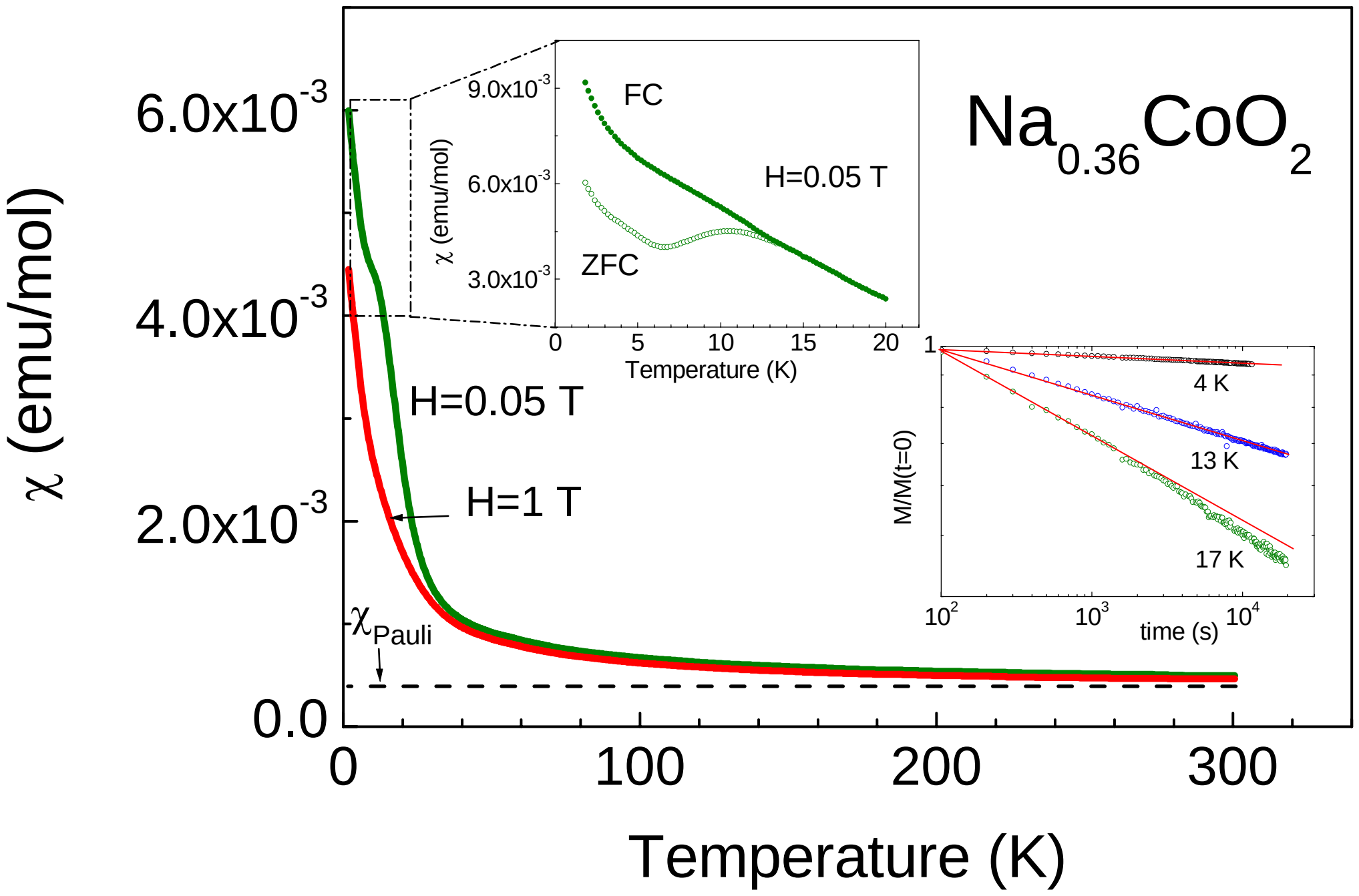

Fig.2- ZFC magnetic susceptibility at two different fields in metallic x=0.36. The dotted line is the Pauli susceptibility determined from high temperature M(H) isotherms. Upper inset: Detail of the ZFC-FC irreversibility visible at low fields below ≈12 K. Lower inset: Relaxation rate of the normalized magnetization. Above the $T_{irr}$ relaxation departs from the slower logarithmic behavior observed at low temperatures.

The spin polarization achieved with an applied magnetic field attenuates the amplitude of the magnetic fluctuations and restores progressively the normal FL behavior. This result is consistent with the observation of an anomalously large $AT^2$ coefficient in the electrical resistivity of single crystals of $Na_{0.7}CoO_2$, which recovers normal values at high fields.[11] It seems clear from our results that previous interpretation of the temperature dependence of $\chi$(T) (in terms of a Curie-Weiss) are incorrect, and could be influenced by a non-homogenous Na distribution in the sample.

In Fig. 2, $\chi$(T) is shown for composition x=0.36 at two different fields. This behavior has been observed in other samples and is representative of compositions around x≈0.3-0.4. A clear irreversibility between the zero-field and field cooling (ZFC-FC) magnetization curves shows up below $T_{irr}$≈12 K, which is suppressed at moderately large fields (not visible already above 0.1 T). The existence of the irreversibility, its field dependence, and the logarithmic variation of the relaxation rate below $T_{irr}$ (Fig. 2, lower inset) would suggest the emergence of a new phase with spin-glass (SG) dynamics at low temperatures and for low values of x. In the SG phase $\chi$(T), remains strongly temperature-dependent down to the lowest temperature probed (1.8 K), not recovering the constant FL behavior. Rather than SG being the cause of the non-FL phase, we believe it is only an effect of the underlying mechanism that introduces strong quasiparticle correlations. One of the most studied origins of the the breakdown of the FL description in itinerant paramagnets is the proximity to a QPT.[12] The experimental signatures of the proximity to a second-order phase transition are the divergent character of the susceptibility, and over all, its scaling. In Fig. 3 we show the divergence of $\chi$(T) as T→0 K, and the progressive recovery of the FL behavior as H increases. Below ≈ 5

K, the low temperature susceptibility shows a power-law temperature dependence of the form

$$\chi = T^{-\eta} \qquad [1]$$

with $\eta$ strongly field dependent. The values of $\eta$ go from 0.30 at low fields to close to zero at high fields, which indicates recovery of the FL state at low temperatures due to the field suppression of spin fluctuations.

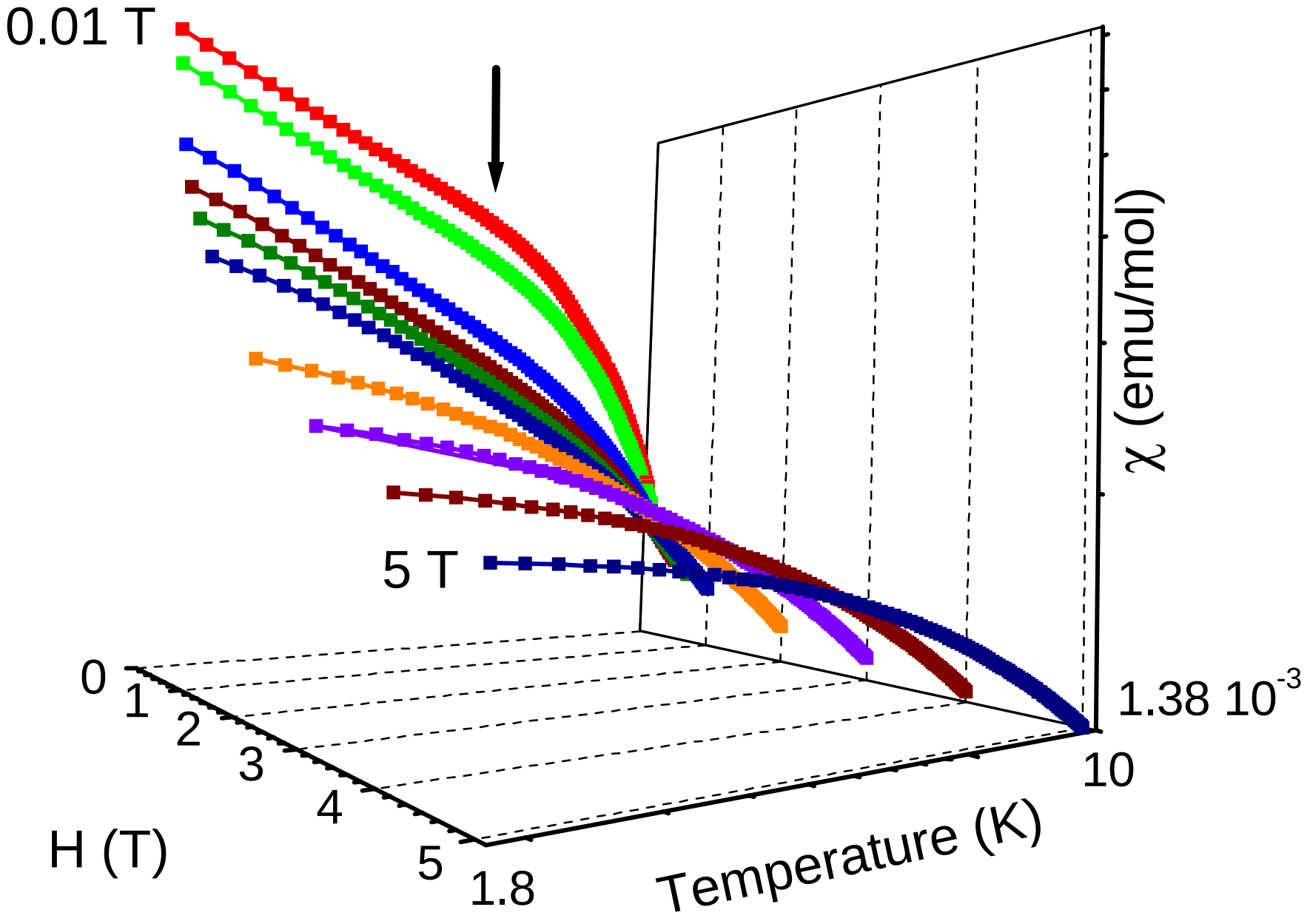

Fig.3- The divergent character of the low field $\chi(T\rightarrow 0)$ fits to equation [1] below ≈ 5 K (marked by the arrow). At high fields the susceptibility recovers FL behaviour, with a decreasing constant value of $\chi(T\rightarrow 0)$ as H increases.

Close to a second order QPT, hyperscaling relationships ensure that the following scaling law must be obeyed by the singular part of the susceptibility once the reduced temperature has been replaced by T to account for the 0 K transition temperature[13,14]

$$\frac{M}{H}T^{\eta} = f\left(\frac{H}{T^{\beta}}\right) \qquad [2]$$

The result of the scaling is shown in Fig. 4. The value of $\eta$ is consistent with the fitting of the low-field susceptibility to equation [1], and both $\eta$ and $\beta$ are internally consistent ($1+\eta/2=\beta$) within the error. For a true FL, no such scaling behavior should be observable as its mere presence shows that there is an energy scale other than Fermi energy that dominates its thermodynamic properties. The exponent $\beta$ being larger than 1 indicates that correlation (extended) fluctuations can be responsible for the singular behavior of $\chi(T\rightarrow 0)$. However, the range of doping explored ($0.36\leq x\leq 0.4$) is not as large as it should be, and no irrefutable conclusion can be derived from these experiments about the localized or extended nature of the spin fluctuations (or even the existence of both types at the same time). The invariance of the critical exponent in a wide range of doping should be tested to discard completely the single-ion origin of the magnetic excitations.

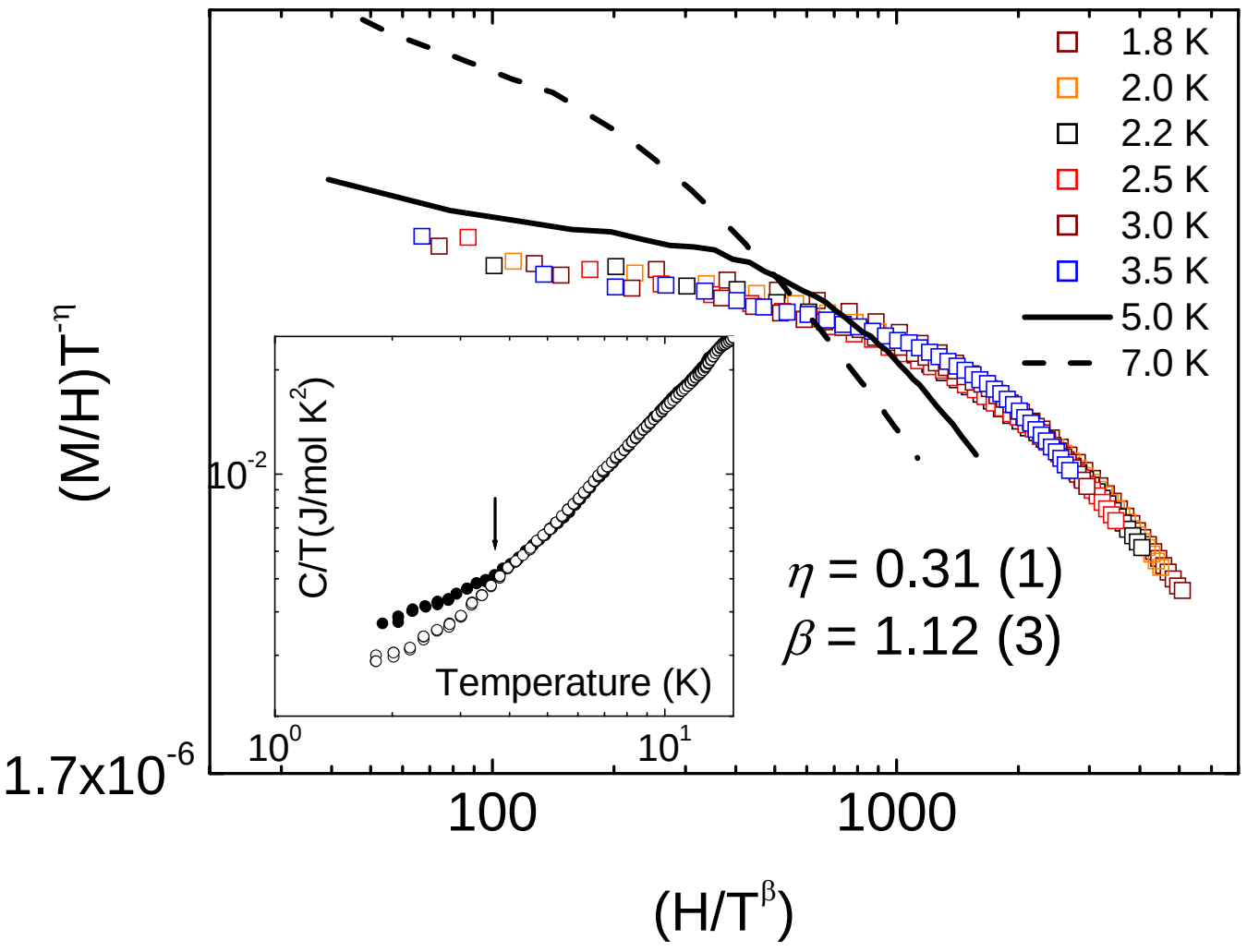

Fig. 4- Scaling of the divergent low temperature susceptibility. The data departs from the low temperature scaling above ≈ 4 K, *i.e.* as the limit of the divergent behavior in $\chi(T\rightarrow 0)$ is approached from below (see Fig. 3). The 5 K and 7 K curves are shown as an example of deviation of the scaling at higher temperatures. Inset: Specific heat data at H=0 (closed symbols) and at H=5T (open symbols).The arrow at 4 K marks the field dependent regime.

A fundamental issue that has to be explored is the origin of the SG phase and the nature of the magnetic fluctuations at low temperature, as well as its relationship with the appearance of superconductivity. $Na_xCoO_2$ is an itinerant paramagnet normally considered as a frustrated antiferromagnet due to the triangular arrangement of the Co atoms in the close-packed $CoO_2$ planes.[15] However, according to Goodenough,[16] a simple extension of the cation-cation superexchange spin correlations for localized electrons to the low-spin octahedral-site Co-$t_{2g}$ itinerant electrons would predict the existence of an itinerant FM ground state for this material. Even partial splitting of the threefold-degenerate $t_{2g}$ orbitals into an $a_1^T$ and twofold-degenerate $e^T$ orbitals by the trigonal field, will keep the same prediction. The O:*2p*-Co:$e_g$ rehybridization process proposed by Marianetti *et al.*,[17] would introduce the possibility of a Zener-like double exchange mechanism, also resulting in a ferromagnetic, metallic ground state like in $SrCoO_3$ or $Sr_2CoO_4$.[18,19] The contraction of the *a*-axis on Na removal, as well as the almost constant cobalt valence,[20] ensures that rehybridization is indeed playing a role in the material. These arguments are in clear agreement with NMR experiments which demonstrate the existence of strong FM fluctuations. Fluctuations can be the source of a short-range partial ordering of the conduction electrons, as it was observed in the high-pressure non-FL phase of MnSi.[21] The coexistence of partially ordered regions (super-spins) with FL electrons, was proposed to be responsible of the non-FL and SG characteristics of many heavy-fermions,[22] in which a divergence of the type $\chi(T) \propto C(T)/T \propto T^{-1+\lambda}$ was predicted.[23] This electronically inhomogeneous state is characterized by an exponent $\lambda<1$, and is equivalent to the *Griffiths* phase of dilute magnetic systems.[24] Reanalysis of our data following this argument leads to $\lambda \approx 0.7$ from the fitting of the divergent low-temperature susceptibility. The results for C(T)/T are shown in the inset of Fig. 4. The low temperature data is not constant, and is strongly field dependent, which indicates the presence of spin fluctuations that are progressively suppressed by the magnetic field. The raw data does not show the characteristic upturn of the *Griffiths* phase at lowest temperature probed. However, when the lattice contribution is subtracted from these data ($C_{lattice}(T) \propto \beta T^3$; $\beta \approx 0.22$ mJ/molK$^4$ obtained from the high-temperature, field-independent part), the electronic contribution of the specific heat diverges as $T^{-1+\lambda}$ ($\lambda \approx 0.6$ increasing with the field). However, the subtraction of the lattice contribution to the specific heat is greatly influenced by the temperature interval fitted. So, $\lambda$ in $C(T)/T \propto T^{-1+\lambda}$ cannot be determined with the accuracy needed to be conclusive about the *Griffiths*-nature of the non-FL phase that develops at low temperature in the metallic precursor of the superconductor.

Based on the analysis of different samples, the (T,H) phase diagrams are presented in Fig. 5 for the large-Na metallic phase and the low-Na metallic precursor of the superconductor. As we move across the series of $Na_xCoO_2$ by changing the Na content, departure from FL behavior appears in the metallic precursor of the superconducting phase. The characteristic temperature below which FL behavior is recovered increases with H, which supports the hypothesis of the FM character of the fluctuations.

The proximity of this material to a QPT makes many competing states become nearly degenerate at low temperatures (the large m* introduces a high density of states at $E_F$) and the systems becomes very susceptible to the formation of new phases, among which superconductivity is a possibility, and has indeed been observed in many materials artificially tuned across or close to a QPT.[6,25]

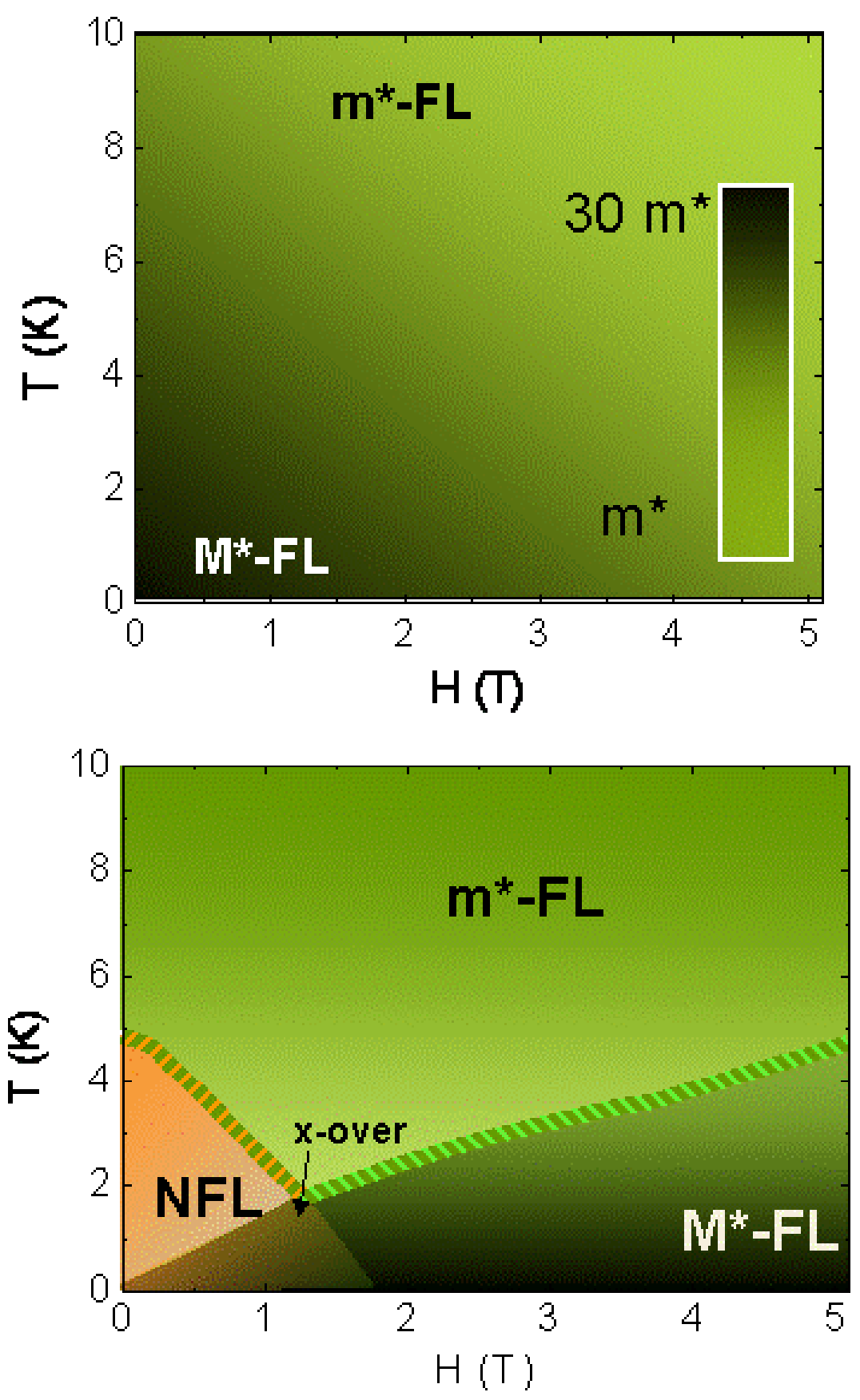

Fig. 5- Temperature vs. magnetic field phase diagram for $Na_xCoO_2$, for high (top) and low (bottom) Na content.

An important ingredient to understand the differences between the low-temperature behavior of the two representative compositions presented in Fig. 5 must be the continuous expansion of the *c*-axis parameter as Na is removed. This reduces the effective electronic dimensionality, which makes the interactions more effective in correlating (destroying) the quasiparticles, which collapse into collective excitations.

In an alternative view of the problem of heavy FL near a QPT, Shaginyan[26] proposed the existence of a new kind of electron liquid state (fermion condensate) which can be reached from an electronically disordered phase (FL) at 0 K across a QPT. When the strongly correlated electron liquid is close to the QCP, its low energy excitations are quasiparticles, but with an m* strongly (T,H) dependent, in contrast to Landau classical quasiparticles. In fact, $m^* \propto H^{-0.5}$ (close to the observed $m^* \propto 1/H^{\approx -0.4}$ for x=0.69), is a possibility predicted in this scenario for a system approaching the QCP from the ordered side. Moreover, when the system is at the critical point itself, m* $\rightarrow\infty$ and so a divergent $\chi(T \rightarrow 0)$ is expected, recovering normal FL behavior at increasing temperatures with H (consistent with the results depicted in Fig. 3 and 5). In the case of $Na_xCoO_2$, dimensionality, doping, and magnetic field must be playing the role of a control parameter that places the system at a distance from the QCP.

In summary, we have shown that metallic $Na_xCoO_2$ is a FL at high doping levels of Na, with the quasiparticle m* becoming heavily renormalized and field dependent at low temperature. When the Na content is around the optimum values for superconductivity, a new phase with spin-glass and non-FL characteristics develops at low temperature. Many experimental evidences suggest that this behavior is a consequence of the proximity to a second-order magnetic QPT. Although measurements at lower temperature should be desirable, previous transport experiments down to 40 mK prove that the system never orders, which together with the scaling of Fig. 4 is a strong argument to support our hypothesis. Our results make superconducting $Na_{0.36}CoO_2 \bullet yH_2O$ a clear candidate for magnetically mediated pairing.

**Acknowledgments**

V. R. Shaginyan, J. B. Goodenough, G. Martinez, J. Merino, and J. Fernández-Rossier are acknowledge for discussion and critical reading of the manuscript. We also thank MEC (Spain) for financial support through MAT2004-05130 and Ramón y Cajal program.

---

[1] K. Takada, H. Sakurai, W. T. Muromachi, F. Izumi, R. A. Dilanian, T. Sasaki, Nature **422**, 53 (2003).

[2] M. L. Foo, Y. Wang, S. Watauchi, H. W. Zandbergen, T. He, R. J. Cava, and N. P. Ong, Phys. Rev. Lett. **92**, 247001 (2004).

[3] F. Rivadulla, J.-S. Zhou, J. B. Goodenough, Phys. Rev. B **68**, 75108 (2003).

[4] Y. Wang, N. S. Rogado, R. J. Cava, and N. P. Ong, Nature **423**, 425 (2003).

[5] D. J. Singh, *Phys. Rev. B* **68,** 20503 (2003).

[6] S. S. Saxena *et al.*, Nature **406**, 587 (2000).

[7] T. Waki. *et al*. J. Phys. Soc. Japan **72**, 3041 (2003); *ibid* **73**, 2963 (2004).

[8] A. T. Boothroyd, R. Coldea, D. A. Tennant, D. Prabhakaran, L. M. Helme, and C. D. Frost, Phys. Rev. Lett. **92**, 197201 (2004).

[9] H. Sakurai, K. Takada, S. Yoshii, T. Sasaki, K. Kindo, E. T. Muromachi, Phys. Rev. B **68**, 132507 (2003).

[10] Y. Ihara, K. Ishida, C. Michioka, M. Kato, K. Yoshimura, H. Sakurai, E. T. Muromachi, J. Phys. Soc. Japan **73**, 2963 (2004).

[11] S. Y. Li *et al*. Phys. Rev. Lett. **93**, 56401 (2004).

[12] S. Sachdev, *Quantum Phase Transitions*, Cambridge University Press, UK (1999).

[13] N. Goldenfeld in "Lectures on phase transitions and the renormalization group", Frontiers in Physics Vol 85, Addison-Wesley, NY 1992.

[14] G. R. Stewart, Rev. Mod. Phys. **73**, 797 (2001).

[15] N. P. Ong and R. J. Cava, Science **305**, 52, (2004).

[16] J. B. Goodenough, in *Magnetism and the Chemical Bond*, John Willey & Sons, New York (1963).

[17] C. A. Marianetti, G. Kotliar, G. Ceder , Phys. Rev. Lett. **92**, 196405 (2004).

[18] R. H. Potze, G. A. Sawatzky, M. Abbate, Phys. Rev. B **51**, 11501 (1995).

[19] J. Matsuno, Y. Okimoto, Z. Fang, X. Z. Yu, Y. Matsui, N. Nagaosa, M. Kawasaki, Y. Tokura, Phys. Rev. Lett. **93**, 167202 (2004).

[20] M. Bañobre-López, F. Rivadulla, R. Caudillo, M. A. López-Quintela, J. Rivas, J. B. Goodenough, Chem. Mater, *in press*.

[21] C. Pfleiderer, D. Reznik, L. Pintschovius, H. v. Löhneysen, M. Garst, A. Rosch, Nature **427**, 227 (2004).

[22] E. D. Bauer et al. Phys. Rev. Lett. **94**, 46401 (2005).

[23] A. H. Castro Neto, G. Castilla, B. A. Jones, Phys. Rev. Lett. **81**, 3531 (1998).

[24] R. B. Griffiths, Phys. Rev. Lett. **23**, 17 (1969).

[25] C. Pfleiderer *et al*. Nature **412**, 58 (2001).

[26] V. R. Shaginyan, J. Exp. Theor. Phys. Lett. **79**, 286 (2004); V. R. Shaginyan *et al.*, cond-mat/0501093.